\documentclass[pra,aps,amsmath,amssymb,twocolumn,showkeys,showpacs]{revtex4}
\usepackage{graphicx}
\newcommand{\hh}{{\mathcal{H}}}

\newcommand{\lnp}{{\mathcal{L}}}
\newcommand{\lsp}{{\mathcal{L}}_{+}}

\newcommand{\niz}{{\mathbf{0}}}
\newcommand{\pen}{\openone}
\newcommand{\rmc}{{\mathrm{C}}}
\newcommand{\rmd}{{\mathrm{D}}}
\newcommand{\rmm}{{\mathrm{M}}}
\newcommand{\Tr}{{\mathrm{Tr}}}
\newcommand{\id}{{\mathrm{id}}}

\newcommand{\ron}{{\mathrm{ran}}}
\newcommand{\bro}{\hat{\rho}}
\newcommand{\vbro}{\hat{\varrho}}
\newcommand{\bsg}{\hat{\sigma}}
\newcommand{\bdl}{\hat{\delta}}
\newcommand{\bom}{\hat{\omega}}
\newcommand{\wbro}{\hat{\varpi}}
\newcommand{\cle}{{\mathcal{E}}}
\newcommand{\cli}{{\mathcal{I}}}

\newcommand{\am}{\hat{A}}
\newcommand{\bn}{\hat{B}}
\newcommand{\km}{\hat{K}}
\newcommand{\wkm}{\hat{L}}
\newcommand{\pqm}{\hat{P}}
\newcommand{\qpm}{\hat{Q}}
\newcommand{\ax}{\hat{X}}
\newcommand{\az}{\hat{Z}}

\begin{document}
\clearpage
\preprint{}

\title{Quantum-coherence quantifiers based on the Tsallis relative $\alpha$-entropies}

\author{Alexey E. Rastegin}
\affiliation{Department of Theoretical Physics, Irkutsk State University,
Gagarin Bv. 20, Irkutsk 664003, Russia}

\begin{abstract}
The concept of coherence is one of cornerstones in physics. The
development of quantum information science has lead to renewed
interest in properly approaching the coherence at the quantum
level. Various measures could be proposed to quantify coherence of
a quantum state with respect to the prescribed orthonormal basis.
To be a proper measure of coherence, each candidate should enjoy
certain properties. It seems that the monotonicity property plays
a crucial role here. Indeed, there is known an intuitive measure
of coherence that does not share this condition. We study
coherence measures induced by quantum divergences of the Tsallis
type. Basic properties of the considered coherence quantifiers are
derived. Trade-off relations between coherence and mixedness are
examined. The property of monotonicity under incoherent selective
measurements has to be reformulated. The proposed formulation can
naturally be treated as a parametric extension of its standard
form. Finally, two coherence measures quadratic in moduli of
matrix elements are compared from the monotonicity viewpoint.
\end{abstract}


\maketitle

\pagenumbering{arabic}
\setcounter{page}{1}

\section{Introduction}\label{sec1}

Interest in the nature of coherent objects and processes in
physics has a very long history. Probably, the notion of coherence
is most known due to the role of phase coherence in optical
phenomena \cite{mandelw}. It is now clear that quantum coherence
is very essential in studying thermodynamic properties of small
systems at low temperatures
\cite{horodecki13,rfag13,rudolph15,rudolx15,ngour15}.
Understanding quantum phenomena such as multipartite entanglement
is also connected with a description of coherence. Entangled
states play a central role in quantum information science
\cite{nielsen}. Due to interaction with the environment, coherent
superpositions of states will be altered. Physical processes
describing decoherence are also the subject of active research.
Recently, many efforts have been made in studies of quantum
coherence as a physical resource. Quantum resource theories are
speciated by a restriction on the quantum operations that can be
implemented \cite{spekkens08,bhors13}. To reveal this question
with respect to coherence, a unified framework for its
quantification is desired. The authors of \cite{bcp14} considered
properties that should be satisfied by any proper measure of
coherence. They also proposed some ways to construct easily
computable measures of coherence. Further development of this
approach was established in \cite{winter15}.

In principle, any measure of distinguishability of quantum states
leads to a candidate for a coherence quantifier \cite{bcp14}. The
following negative result should be emphasized here. It turns out
that the measure induced by the squared Hilbert-Schmidt norm does
not enjoy a valid coherence monotonicity \cite{bcp14}. In this
regard, monotonicity properties play the crucial role in
development of proper coherence measures. The coherence
quantifiers of \cite{bcp14} were used for obtaining 
complementarity relations for quantum coherence with respect to
mutually unbiased bases \cite{hall15,mondal15}. The authors of
\cite{hall15} also claimed a conjecture related to the negative
result mentioned above. Due to a simple structure, the conjectured
quantifier of coherence deserves further investigations. Relations
between coherence and multi-path interference phenomena were
considered in \cite{bera15,bagan15}. The role of coherence in the
Deutsch--Jozsa and related algorithms is considered in
\cite{hillery15}. The authors of \cite{bca15} examined under which
conditions the coherence of an open quantum system is unaffected
by noise. The paper of \cite{vedral15} is devoted to quantum
processes that can neither create nor use coherence.
Quantification of coherence in infinite-dimensional systems is
studied in \cite{jxu15,zslf16}.

In this work, we study coherence quantifiers based on the Tsallis
relative entropies. Quantum relative entropies of the Tsallis
type are expressed in terms of powers of density matrices. Hence,
we may expect a relatively simple character of induced coherence
measures. The paper is organized as follows. In Sect. \ref{sec2},
we recall the approach developed in \cite{bcp14} and list some
preliminaries. In Sect. \ref{sec3}, we consider relative entropies
of the Tsallis type and prove the two results required. In Sect.
\ref{sec4}, we study properties of coherence measures based on the
Tsallis relative entropies. In particular, trade-off relations
between coherence and mixedness are obtained. The case of a single
qubit is separately discussed in Sect. \ref{sec5}. The
monotonicity property is satisfied with an interesting form found
in Sect. \ref{sec6}. The obtained family of coherence quantifiers
includes a homogeneous quadratic function of moduli of matrix
elements. Another quadratic function of such a kind is induced by
the squared $\ell_{2}$-norm. In Sect. \ref{sec7}, the two
quadratic measures of coherence are compared within a concrete
example. In Sect. \ref{sec8}, we conclude the paper with a summary
of results.

\section{Preliminaries}\label{sec2}

In this section, we briefly recall basic points of the approach of
\cite{bcp14} to quantifying quantum coherence. In principle,
measures of coherence may be introduced with using operator norms.
Some genuine properties of coherence measures are related to their
behavior with respect to quantum operations. Thus, main results of
quantum operation formalism should be used. Let $\lnp(\hh)$ be the
space of linear operators on finite-dimensional Hilbert space
$\hh$. By $\lsp(\hh)$, we denote the set of positive semidefinite
operators on $\hh$. By $\ron(\ax)$, we denote the range of
operator $\ax$. In the following, we use the convention that
powers of a positive operator are taken only on its support. For
any $\az\in\lsp(\hh)$, we treat $\az^{0}$ as the orthogonal
projector onto $\ron(\az)$. Let $\pqm$ and $\qpm$ be operators of
the orthogonal projection. In the finite-dimensional case, we
define $\pqm\vee\qpm$ as the projector onto the sum of subspaces
$\ron(\pqm)+\ron(\qpm)$. In the infinite-dimensional case, this
definition should be modified. In the following, we will deal with
finite dimensions only.

A distance between operators of interest can be characterized by
norms. With respect to the given orthonormal basis, each operator
$\ax:{\>}\hh\rightarrow\hh$ is represented by the square matrix
with elements $x_{ij}$. The $\ell_{1}$-norm is then defined
as \cite{hornJ}
\begin{equation}
\|\ax\|_{\ell_{1}}:=\sum\nolimits_{ij} |x_{ij}|
\, . \label{ell1n}
\end{equation}
Further, the $\ell_{2}$-norm is defined as
\begin{equation}
\|\ax\|_{\ell_{2}}:=
\left(
\sum\nolimits_{ij} |x_{ij}|^{2}
\right)^{\!1/2}
. \label{ell2n}
\end{equation}
This norm is also known as the Frobenius or Hilbert--Schmidt norm
\cite{hornJ}. There are other frequently used norms such as the
Schatten norms and the Ky Fan norms. These norms, defined in terms
of singular values, are unitarily invariant \cite{hornJ}.

A state of the quantum system of interest is represented by
positive operator $\bro$ normalized as $\Tr(\bro)=1$. To formulate
the desired properties of measures of coherence, we will use some
basic facts about quantum operations. Let us consider a linear map
\begin{equation}
\Phi:{\>}\lnp(\hh_{A})\rightarrow\lnp(\hh_{B})
\, , \label{hdhdp}
\end{equation}
where the input space $\hh_{A}$ and the output space $\hh_{B}$ can
be different. This map is positive, when
$\Phi(\ax)\in\lsp(\hh_{B})$ for each $\ax\in\lsp(\hh_{A})$
\cite{nielsen}. Physical processes are described by completely
positive maps \cite{nielsen}. Let $\id_{R}$ be the identity map on
$\lnp(\hh_{R})$, where the Hilbert space $\hh_{R}$ is related to
an imagined reference system. The complete positivity implies that
the map $\Phi\otimes\id_{R}$ is positive for arbitrary
dimensionality of $\hh_{R}$. Each completely positive map can be
represented in the form \cite{nielsen}
\begin{equation}
\Phi(\ax)=\sum\nolimits_{n}\km_{n}\ax\km_{n}^{\dagger}
\, , \label{osrp}
\end{equation}
with the Kraus operators $\km_{n}:{\>}\hh_{A}\rightarrow\hh_{B}$.
The map preserves the trace, when these operators satisfy
\begin{equation}
\sum\nolimits_{n}\km_{n}^{\dagger}\km_{n}=\pen_{A}
\, . \label{clrl}
\end{equation}
Trace-preserving completely positive (TPCP) maps are usually
referred to as quantum channels.

The authors of \cite{bcp14} developed an approach to quantum
coherence with the use of a fixed preferred basis for a physical
situation of interest. They also collected desirable properties a
proper measure of coherence should satisfy. Some applications of
these ideas were further developed in \cite{hall15,bagan15,jxu15}.
Let $\cle=\bigl\{|e_{i}\rangle\bigr\}$ be a prescribed orthonormal
basis in $\hh_{A}$. The set of incoherent states contains all
states that are diagonal with respect to $\cle$, namely,
\begin{equation}
\bdl=\sum\nolimits_{i} \delta_{i}\,|e_{i}\rangle\langle{e}_{i}|
\, . \label{incs}
\end{equation}
By $\cli(\cle)\subset\lsp(\hh_{A})$, we mean the set of all such
states. Quantifiers of coherence should map from the set of states
to the set of non-negative real numbers. The following two
quantifiers of coherence are intuitively natural \cite{bcp14}.
Using the $\ell_{1}$-norm finally gives
\begin{equation}
\rmc_{\ell_{1}}(\cle|\bro):=\underset{\bdl\in\cli(\cle)}{\min}\bigl\|\bro-\bdl\bigr\|_{\ell_{1}}=
\sum_{i\neq{j}}\bigl|\langle{e}_{i}|\bro|e_{j}\rangle\bigr|
\, . \label{c1nrm}
\end{equation}
Another natural candidate is the one based on the squared
$\ell_{2}$-norm \cite{bcp14}. That is, we write
\begin{equation}
\rmc_{\ell_{2}}(\cle|\bro):=\underset{\bdl\in\cli(\cle)}{\min}\bigl\|\bro-\bdl\bigr\|_{\ell_{2}}^{2}=
\sum_{i\neq{j}}\bigl|\langle{e}_{i}|\bro|e_{j}\rangle\bigr|^{2}
\, . \label{c22nrm}
\end{equation}
Unfortunately, this seemingly natural measure does not obey the
monotonicity requirement \cite{bcp14}. The trace norm also induces
an interesting candidate for quantification of coherence
\cite{rpl15}. The authors of \cite{ssdba15} proposed a common
frame to quantify quantumness in terms of coherence and
entanglement. They also derived the geometric measure of coherence
based on the notion of fidelity of quantum states
\cite{uhlmann76,jozsa94}.

\section{Quantum divergences of the Tsallis type}\label{sec3}

In this section, we recall the definition of relative entropies of
the Tsallis type. Many fundamental results of quantum information
theory are connected with the properties of the standard relative
entropy \cite{nielsen}. There exist several extensions of the
standard entropic functions \cite{bengtsson}. Many quantum
relative entropies can be unified within the concept of
$f$-divergences \cite{hmpb11}. This approach is a quantum
counterpart of the Csisz\'{a}r $f$-divergence \cite{ics67}. For
$0<\alpha\neq1$, the Tsallis relative $\alpha$-entropy is defined
as \cite{borland,fky04}
\begin{equation}
D_{\alpha}(p||q):=
\frac{1}{\alpha-1}
\left(
\sum\nolimits_{j} p_{j}^{\alpha}q_{j}^{1-\alpha}-1
\right)
. \label{tfdf}
\end{equation}
If for some $j$ we have $p_{j}\neq0$ and $q_{j}=0$, and then the
relative $\alpha$-entropy with $\alpha>1$ is set to be $+\infty$.
In the limit $\alpha\to1$, the above divergence gives the standard
relative entropy $D_{1}(p||q)=\sum_{j}p_{j}\,\ln(p_{j}/q_{j})$.
Here, one assumes $-0\ln0\equiv0$ and $-p_{j}\ln0\equiv+\infty$
for $p_{j}>0$ \cite{nielsen}. Basic properties of quantity (\ref{tfdf})
were discussed in \cite{borland,fky04}. We mention only
several of them. It was shown in \cite{fky04}
$D_{\alpha}(p||q)\geq0$. Necessary conditions for vanishing
$D_{\alpha}(p||q)$ are a more complicated question. For the class
of Csisz\'{a}r $f$-divergences, this question was considered in
\cite{vajda06}. The answer is connected with the notion of strict
convexity of a certain function at $1$. It follows from the results
of \cite{vajda06} that $D_{\alpha}(p||q)=0$ only when the
distributions $p$ and $q$ coincides (see, e.g., example 2 of
\cite{vajda06}).

We shall now recall the notion of quantum relative entropy. For
density operators $\bro$ and $\bsg$, the quantum relative entropy
is expressed as \cite{nielsen}
\begin{equation}
\rmd_{1}(\bro||\bsg):=
\begin{cases}
\Tr(\bro\,\ln\bro-\bro\,\ln\bsg) \,,
& \text{if $\ron(\bro)\subseteq\ron(\bsg)$} \,, \\
+\infty\, , & \text{otherwise} \,.
\end{cases}
\label{relan}
\end{equation}
For a discussion of the role of (\ref{relan}) in quantum
information theory, see \cite{nielsen,vedral02} and references
therein. The divergence (\ref{relan}) can be generalized in
several ways. For $\alpha\in(1;+\infty)$, the Tsallis
$\alpha$-divergence is defined as
\begin{equation}
\rmd_{\alpha}(\bro||\bsg):=
\begin{cases}
\frac{\Tr(\bro^{\alpha}\bsg^{1-\alpha})-1}{\alpha-1} \, ,
& \text{if $\ron(\bro)\subseteq\ron(\bsg)$} \,, \\
+\infty\, , & \text{otherwise} \,.
\end{cases}
\label{qendf}
\end{equation}
When $\ron(\bro)\subseteq\ron(\bsg)$, the trace is assumed to be
taken over $\ron(\bsg)$. For $\alpha\in(0;1)$, the first entry can
be used without such conditions. The formula (\ref{qendf}) can be
represented similarly to (\ref{relan}) with the use of the
$\alpha$-logarithm. For $0<\alpha\neq1$ and real $\xi>0$, the
$\alpha$-logarithm is defined as \cite{borland}
\begin{equation}
\ln_{\alpha}(\xi):=\frac{\xi^{1-\alpha}-1}{1-\alpha} \ .
\label{alog}
\end{equation}
For $\alpha\to1$, the function (\ref{alog}) reduces to the usual
logarithm. Up to a factor, the relative entropy (\ref{qendf}) is a
particular case of quasi-entropies proposed by Petz \cite{petz86}.
A more general family of quantum $f$-divergences is studied in
\cite{hmpb11}. The following extension will also be useful. Let
$\am$ and $\bn$ be positive operators such that
$\ron(\am)\subseteq\ron(\bn)$. The $\alpha$-divergence of $\am$
with respect to $\bn$ is defined by
\begin{equation}
\rmd_{\alpha}(\am||\bn):=
\frac{1}{\alpha-1}\Bigl[\Tr(\am^{\alpha}\bn^{1-\alpha})-\Tr(\am)\Bigr]
. \label{dfab}
\end{equation}
Recall several properties of the quantum $\alpha$-divergence. They
follows from the corresponding results on the quantum
$f$-divergences \cite{hmpb11}. For all $\lambda\in[0;+\infty)$, we
have
\begin{equation}
\rmd_{\alpha}(\lambda\am||\lambda\bn)=\lambda\,\rmd_{\alpha}(\am||\bn)
\, . \label{flam}
\end{equation}
Let four positive semidefinite operators $\am_{1}$, $\bn_{1}$, $\am_{2}$,
$\bn_{2}$ obey $\am_{1}^{0}\vee\bn_{1}^{0}\perp\am_{2}^{0}\vee\bn_{2}^{0}$;
then
\begin{equation}
\rmd_{\alpha}\bigl(\am_{1}+\am_{2}\big|\big|\bn_{1}+\bn_{2}\bigr)
=\rmd_{\alpha}(\am_{1}||\bn_{1})+\rmd_{\alpha}(\am_{2}||\bn_{2})
\, . \label{tdit}
\end{equation}
The latter can be proved for quantum $f$-divergences under certain
conditions \cite{hmpb11}. We will use (\ref{tdit}) in studies of the
monotonicity of coherence quantifiers.

One of the fundamental properties of (\ref{relan}) is its
monotonicity under TPCP maps \cite{nielsen}. That is, for any TPCP
map (\ref{hdhdp}) we have
$\rmd_{1}\bigl(\Phi(\bro)\big|\big|\Phi(\bsg)\bigr)\leq\rmd_{1}(\bro||\bsg)$
In the classical regime, the relative Tsallis entropy (\ref{tfdf}) is
monotone under stochastic maps for all $\alpha\geq0$ \cite{fky04}.
This is not the case for the quantum regime. The quantum
$\alpha$-divergence (\ref{qendf}) is monotone under TPCP maps for
$\alpha\in(0;2]$, namely,
\begin{equation}
\rmd_{\alpha}\bigl(\Phi(\bro)\big|\big|\Phi(\bsg)\bigr)\leq\rmd_{\alpha}(\bro||\bsg)
\, . \label{mnren}
\end{equation}
This claim follows from the results of \cite{hmpb11} (see theorem
4.3 therein) and the two facts about functions on positive
matrices. The function $\xi\mapsto\xi^{\alpha}$ is matrix concave
on $[0;+\infty)$ for $0\leq\alpha\leq1$ and matrix convex on
$[0;+\infty)$ for $1\leq\alpha\leq2$ (see, e.g., theorems 4.2.3
and 1.5.8 in \cite{bhatia07}). The monotonicity further yields the
joint convexity of the $f$-divergences (see, e.g., corollary 4.7
of \cite{hmpb11}). In particular, the quantum $\alpha$-divergences
of the Tsallis type are jointly convex for $\alpha\in(0;2]$. Let
$\{\bro_{n}\}$ and $\{\bsg_{n}\}$ be two collections of density
matrices, and let $p_{n}$'s be positive numbers that are
summarized to $1$. For $\alpha\in(0;2]$, we then have
\begin{equation}
\rmd_{\alpha}\biggl(\sum_{n}p_{n}\bro_{n}\bigg|\bigg|\sum_{n}p_{n}\bsg_{n}\biggr)
\leq\sum_{n}p_{n}\,\rmd_{\alpha}(\bro_{n}||\bsg_{n})
\, . \label{joico}
\end{equation}
The properties (\ref{mnren}) and (\ref{joico}) will be very
important in the verification of corresponding properties of
induced coherence measures. We shall also discuss other properties
of the quantum $\alpha$-divergences. They are essential from the
viewpoint of constructing measures of coherence. So, we present
them as separate statements.

\newtheorem{prop31}{Theorem}
\begin{prop31}\label{thm31}
For $\alpha>0$, the quantum $\alpha$-divergence is non-negative,
\begin{equation}
\rmd_{\alpha}(\bro||\bsg)\geq0
\, , \label{tkln}
\end{equation}
with equality if and only if $\bro=\bsg$.
\end{prop31}

The proof of this statement is carried out similarly to the case
of the standard relative entropy (see, e.g., theorem 11.7 in
\cite{nielsen}). We refrain from presenting the details here. It
should be noted that positivity of the Tsallis $\alpha$-divergence
{\it per se} was considered in proposition 2.4 of \cite{fky04}.
Although the authors of \cite{fky04} focused on the range
$0<\alpha<1$, their arguments are applicable for all positive
$\alpha$. We are also interested in conditions for equality. For
this aim, we merely modify the proof of theorem 11.7 of
\cite{nielsen} with the $\alpha$-logarithm (\ref{alog}). Another
property of the Tsallis $\alpha$-divergence is essential in
studying the monotonicity of the induced coherence measures.

\newtheorem{prop32}[prop31]{Theorem}
\begin{prop32}\label{thm32}
Let $\{\km_{n}\}$ be a set of operators such that
$\sum_{n}\km_{n}^{\dagger}\km_{n}={\textup{\pen}}_{A}$. With the
given normalized density matrices $\bro$ and $\bsg$ on $\hh_{A}$,
one associates two probability distributions with the
corresponding probabilities
\begin{equation}
p_{n}=\Tr(\km_{n}\bro\km_{n}^{\dagger})
\, , \qquad
q_{n}=\Tr(\km_{n}\bsg\km_{n}^{\dagger})
\, . \label{twpr}
\end{equation}
For $\alpha>0$, the quantum $\alpha$-divergences obey
\begin{equation}
\sum_{n}
\rmd_{\alpha}\bigl(\km_{n}\bro\km_{n}^{\dagger}\big|\big|\km_{n}\bsg\km_{n}^{\dagger}\bigr)
\geq
\sum_{n}
p_{n}^{\alpha}q_{n}^{1-\alpha}\,\rmd_{\alpha}(\bro_{n}||\bsg_{n})
\, , \label{twpr1}
\end{equation}
where the states
$\bro_{n}=p_{n}^{-1}\km_{n}\bro\km_{n}^{\dagger}$ and
$\bsg_{n}=q_{n}^{-1}\km_{n}\bsg\km_{n}^{\dagger}$ are
normalized.
\end{prop32}

{\bf Proof.} The right-hand side of (\ref{twpr1}) is focused on
those values of $n$ that $p_{n}\neq0$ and $q_{n}\neq0$
simultaneously. When $p_{n}\neq0$ and $q_{n}=0$ for some $n$, we
have $\km_{n}\bro\km_{n}^{\dagger}\neq\niz$ and
$\km_{n}\bsg\km_{n}^{\dagger}=\niz$. Then the corresponding term
in the left-hand side of (\ref{twpr1}) becomes $+\infty$
[see the second line of (\ref{qendf})], whence the statement
holds. So, we will prove (\ref{twpr1}) for the case, in which
$p_{n}\neq0$ implies $q_{n}\neq0$. Due to the definition
(\ref{qendf}), we can write
\begin{align}
&\rmd_{\alpha}\bigl(\km_{n}\bro\km_{n}^{\dagger}\big|\big|\km_{n}\bsg\km_{n}^{\dagger}\bigr)=
\rmd_{\alpha}(p_{n}\bro_{n}||q_{n}\bsg_{n})
\nonumber\\
&=p_{n}^{\alpha}q_{n}^{1-\alpha}\,\rmd_{\alpha}(\bro_{n}||\bsg_{n})+
\frac{p_{n}^{\alpha}q_{n}^{1-\alpha}-p_{n}}{\alpha-1}
\ . \label{dfqen}
\end{align}
Hence, the left-hand side of (\ref{twpr1}) minus the right-hand
side is equal to $D_{\alpha}(p||q)\geq0$. $\blacksquare$

In the case $\alpha=1$, the inequality (\ref{twpr1}) is reduced to
\begin{equation}
\sum_{n}
\rmd_{1}\bigl(\km_{n}\bro\km_{n}^{\dagger}\big|\big|\km_{n}\bsg\km_{n}^{\dagger}\bigr)
\geq
\sum_{n}
p_{n}\,\rmd_{1}(\bro_{n}||\bsg_{n})
\, . \label{twpr11}
\end{equation}
This property of the standard relative entropy was formulated and
proved in \cite{vepl98} [see item (F4) therein]. So, we obtained
an extension of the formula (\ref{twpr11}) to quantum divergences
of the Tsallis type. Such an extension does not seem to have been
previously recognized in the literature. It should be noted,
however, that the right-hand side of (\ref{twpr1}) is more
complicated in character. We will use (\ref{twpr1}) in studying
changes of coherence quantifiers under incoherent selective
measurements.

\section{Coherence quantifiers based on the Tsallis divergences}\label{sec4}

The authors of \cite{bcp14} pointed out a general way to obtain
candidates for quantification of coherence. To find more coherence
measures, we can try to consider generalized relative entropies.
This is formally posed as follows. Let us pick the Tsallis
$\alpha$-divergence as a distinguishability measure. For
$\alpha>0$, we define the quantity
\begin{equation}
\rmc_{\alpha}(\cle|\bro):=
\underset{\bdl\in\cli(\cle)}{\min}\rmd_{\alpha}(\bro||\bdl)
\, . \label{c1dt}
\end{equation}
In principle, this approach could be used with divergences of a more
general type. The optimization problem is generally not easy.
However, it is simply resolved in the case of Tsallis divergences.
The following statement takes place.

\newtheorem{prop41}[prop31]{Theorem}
\begin{prop41}\label{thm41}
For all $0<\alpha\neq1$, the corresponding coherence measure
is expressed as
\begin{equation}
\rmc_{\alpha}(\cle|\bro)=
\frac{1}{\alpha-1}
\left\{
\Bigl(
\sum\nolimits_{j}\langle{e}_{j}|\bro^{\alpha}|e_{j}\rangle^{1/\alpha}
\Bigr)^{\!\alpha}
-1
\right\}
. \label{res1}
\end{equation}
\end{prop41}

{\bf Proof.} Since the $\alpha$-divergence
$\rmd_{\alpha}(\bro||\bdl)$ should be minimized, we will assume
$\ron(\bro)\subseteq\ron(\bdl)$. If $\delta_{j}$'s are eigenvalues
of $\bdl$, we set $\delta_{j}=0$ whenever
$\langle{e}_{j}|\bro|e_{j}\rangle=0$. As any
$\bdl\in\cli(\cle)$ is diagonal with respect to $\cle$, we write
\begin{equation}
\rmd_{\alpha}(\bro||\bdl)=\frac{1}{\alpha-1}
\left\{
\sum\nolimits_{j}\langle{e}_{j}|\bro^{\alpha}|e_{j}\rangle\,\delta_{j}^{1-\alpha}
-1\right\}
, \label{resp1}
\end{equation}
where the sum is taken over non-zero matrix elements. Let us
define the probabilities $r_{j}$ such that
$r_{j}^{\alpha}\propto\langle{e}_{j}|\bro^{\alpha}|e_{j}\rangle$.
Together with the normalization condition, one gets
\begin{align}
&r_{j}=
\frac{1}{N}
\>\langle{e}_{j}|\bro^{\alpha}|e_{j}\rangle^{1/\alpha}
, \label{resp2}\\
&N=\sum\nolimits_{i}\langle{e}_{i}|\bro^{\alpha}|e_{i}\rangle^{1/\alpha}
. \label{resp2n}
\end{align}
Substituting
$\langle{e}_{j}|\bro^{\alpha}|e_{j}\rangle=N^{\alpha}r_{j}^{\alpha}$
into (\ref{resp1}), we obtain
\begin{equation}
\rmd_{\alpha}(\bro||\bdl)=
N^{\alpha}D_{\alpha}(r||\delta)
+\frac{N^{\alpha}-1}{\alpha-1}
\ . \label{resp3}
\end{equation}
Here, the probabilities $r_{j}$ and the normalization denominator
$N$ depends only on the state $\bro$. As was already mentioned, we
always have $D_{\alpha}(r||\delta)\geq0$. To minimize the
right-hand side of (\ref{resp3}), we should therefore reach
$D_{\alpha}(r||\delta)=0$ by setting $\delta_{j}=r_{j}$. Combining
the result with the formula for $N$ completes the proof.
$\blacksquare$

Thus, the result of minimizing is expressed in terms of matrix
elements of the power $\bro^{\alpha}$. For the given $\bro$ and
$\alpha$, the minimum in (\ref{c1dt}) is reached with the state
\begin{equation}
\bdl_{\rho}=\frac{1}{N}
\sum\nolimits_{j}
\langle{e}_{j}|\bro^{\alpha}|e_{j}\rangle^{1/\alpha}
\,|e_{j}\rangle\langle{e}_{j}|
\, . \label{bdlra}
\end{equation}
We avoid stating explicitly that the $\hat{\delta}_{\rho}$ depends on the
value of $\alpha$. It the case $\alpha=1$, we obtain
the formulation with the standard relative entropy. Then
the state (\ref{bdlra}) is obtained from $\bro$ by deleting all
off-diagonal elements. Then the coherence measure is merely equal
to the von Neumann entropy of this diagonal state minus the von
Neumann entropy of $\bro$ \cite{bcp14}. Let us consider another
interesting case $\alpha=2$. As the density matrix is
Hermitian, we obtain
\begin{equation}
\rmc_{2}(\cle|\bro)=
\left(
\sum\nolimits_{j}
\sqrt{\sum\nolimits_{i}|\rho_{ij}|^{2}}
\ \right)^{\!2}
-1
\, , \label{cma2}
\end{equation}
where $\rho_{ij}=\langle{e}_{i}|\bro|e_{j}\rangle$.
This coherence measure is a function of squared moduli
$|\rho_{ij}|^{2}$, but more complicated in comparison with
(\ref{c22nrm}). We now consider basic properties of the
presented coherence quantifiers.

First of all, the quantity (\ref{c1dt}) is zero for all incoherent
states. It follows from (\ref{res1}) by substituting
$\langle{e}_{j}|\bro^{\alpha}|e_{j}\rangle=\rho_{jj}^{\alpha}$ and
$\sum_{j}\rho_{jj}=1$. Further, we have
$\rmc_{\alpha}(\cle|\bro)=0$ only for incoherent states. According
to Theorem \ref{thm31}, $\rmd_{\alpha}(\bro||\bsg)$ is zero only
for $\bro=\bsg$. So, for $\bro\notin\cli(\cle)$ and any
$\bdl\in\cli(\cle)$ we have $\rmd_{\alpha}(\bro||\bdl)>0$. Thus,
the coherence measure (\ref{c1dt}) satisfies one of the conditions
listed in \cite{bcp14}. An upper bound on the $\alpha$-quantifiers
of coherence can be expressed in terms of the purity.

\newtheorem{prop42}[prop31]{Theorem}
\begin{prop42}\label{thm42}
For $0<\alpha\leq2$, we have
\begin{equation}
\rmc_{\alpha}(\cle|\bro)
\leq-\ln_{\alpha}\!\left(
\frac{1}{d\,\Tr(\bro^{2})}
\right)
. \label{res421}
\end{equation}
For $2<\alpha<\infty$, we have
\begin{align}
&\rmc_{\alpha}(\cle|\bro)\leq
\label{res422}\\
&
\frac{1}{\alpha-1}\left\{
d\,\Tr(\bro^{2})
\!\left(
1+\sqrt{d-1}\sqrt{d\,\Tr(\bro^{2})-1}
\,\right)^{\!\alpha-2}-1
\right\}
. \nonumber
\end{align}
\end{prop42}

{\bf Proof.} We first consider the case $\alpha\neq1$. Due to
(\ref{c1dt}), for $0<\alpha\neq1$ we immediately obtain
\begin{equation}
\rmc_{\alpha}(\cle|\bro)\leq\frac{d^{\alpha-1}\Tr(\bro^{\alpha})-1}{\alpha-1}
\ . \label{ubcma}
\end{equation}
The right-hand side of (\ref{ubcma}) is the $\alpha$-divergence of
$\bro$ with respect to the completely mixed state. For
$\alpha\leq2$, the function $\xi\mapsto\xi^{\alpha-1}/(\alpha-1)$
is concave. Calculating traces in the eigenbasis of $\bro$ and
applying Jensen's inequality, we then have
\begin{equation}
\frac{\Tr(\bro^{\alpha})}{\alpha-1}=\sum\nolimits_{j}\lambda_{j}\>\frac{\lambda_{j}^{\alpha-1}}{\alpha-1}\leq
\frac{\bigl[\Tr(\bro^{2})\bigr]^{\alpha-1}}{\alpha-1}
\ . \label{jencor}
\end{equation}
Here, the eigenvalues $\lambda_{j}$ of $\bro$ obey the
normalization condition. Combining (\ref{jencor}) with
(\ref{ubcma}) and (\ref{alog}) finally gives (\ref{res421}) for
$\alpha\neq1$. To complete the proof of (\ref{res421}), we write
\begin{equation}
\rmc_{1}(\cle|\bro)\leq\ln{d}+\Tr(\bro\ln\bro)
\, , \label{ubcma1}
\end{equation}
and repeat the above reasons with the concave function
$\xi\mapsto\ln\xi$.

Let us proceed to the case $\alpha>2$. As follows from lemma 3 of
\cite{rast13b}, the maximal eigenvalue of $\bro$ satisfies
\begin{equation}
\lambda_{\max}\leq\frac{1}{d}
\left(
1+\sqrt{d-1}\sqrt{d\,\Tr(\bro^{2})-1}
\,\right)
. \label{lamax}
\end{equation}
Combining this with
$\Tr(\bro^{\alpha})\leq\lambda_{\max}^{\alpha-2}\,\Tr(\bro^{2})$
and (\ref{ubcma}) completes the proof. $\blacksquare$

The results (\ref{res421}) and (\ref{res422}) provide an upper
bound on the coherence quantifiers in terms of the purity
$\Tr(\bro^{2})$. They are similar to the complementarity relation
derived in \cite{hall15} with the coherence measure (\ref{c1nrm})
taken for $d+1$ mutually unbiased bases (MUBs). The distinction of
the formulas (\ref{res421}) and (\ref{res422}) is that only a
single quantifier is involved. The purity is closely related to
the Brukner--Zeilinger concept of operationally invariant measure
of information in quantum measurements \cite{brz99}. The method of
\cite{brz99} is based on the use of a complete set of $d+1$ MUBs.
Except for prime power $d$, the existence of such sets is an open
problem \cite{bz10}. Then three other schemes to approach the
Brukner--Zeilinger information can be used \cite{rastproca}.
Hence, we have a way to estimate the right-hand sides of both
(\ref{res421}) and (\ref{res422}) in experiment.

The result (\ref{res421}) can be reformulated as a trade-off
relation between coherence and mixedness. For $d$-dimensional
state $\bro$, one of the natural quantifiers of the mixedness is given
by \cite{kwiat2004}
\begin{equation}
\rmm(\bro):=\frac{d}{d-1}\left[1-\Tr(\bro^{2})\right]
. \label{mixdf}
\end{equation}
This figure is zero for pure states and reaches $1$ for the
completely mixed state. The purity can be expressed via the
mixedness and then substituted to (\ref{res421}) and
(\ref{res422}). However, the resulting inequalities will be too
complicated. A convenient method is to approach the right-hand
side of (\ref{res421}) from above by a linear function of the
variable $d\,\Tr(\bro^{2})=y\in[1;d]$. Here, we deal with the
function
\begin{equation}
f_{\alpha}(y):=
-\ln_{\alpha}\!\left(
\frac{1}{y}
\right)
=\frac{y^{\alpha-1}-1}{\alpha-1}
\ , \label{funy}
\end{equation}
which is concave for $\alpha\leq2$. By the Taylor formula with
remainder written in Lagrange's form, with $1<c<d$, one gets
\begin{align}
f_{\alpha}(y)&=
f_{\alpha}(1)+f_{\alpha}^{\prime}(1){\,}(y-1)
+\frac{1}{2}{\>}f_{\alpha}^{\prime\prime}(c){\,}(y-1)^{2}
\nonumber\\
&\leq{y}-1
\, . \label{tay}
\end{align}
Here, we used $f_{\alpha}(1)=0$, $f_{\alpha}^{\prime}(1)=1$, and
$f_{\alpha}^{\prime\prime}(c)\leq0$. The claim (\ref{tay}) poses
that the graph of concave $f_{\alpha}(y)$ goes under its tangent
line drawn at the point $y=1$. Combining (\ref{tay}) with
(\ref{res421}), for $0<\alpha\leq2$ we have
\begin{equation}
\rmc_{\alpha}(\cle|\bro)\leq{d}\,\Tr(\bro^{2})-1
\, . \label{ubcma2}
\end{equation}
Due to (\ref{ubcma2}), we obtain a trade-off relation between
coherence and mixedness in the form
\begin{equation}
\frac{1}{d-1}\>\rmc_{\alpha}(\cle|\bro)+\rmm(\bro)\leq 1
\, , \label{mixcr}
\end{equation}
where $0<\alpha\leq2$. When a degree of mixedness increases, an
upper bound on values of the coherence $\alpha$-quantifier
decreases. It is instructive to compare (\ref{mixcr}) with theorem
1 of the paper \cite{sbdp15}, where trade-off between coherence
and mixedness is expressed in terms of the quantities
(\ref{c1nrm}) and (\ref{mixdf}).

\section{On coherence of a single qubit}\label{sec5}

In this section, we examine coherence of a single qubit with the
use of the $\alpha$-quantifiers. Here, we can express results more
explicitly. With respect to the prescribed basis, the
density matrix is written as
\begin{equation}
\bom=
\begin{pmatrix}
u & w^{*} \\
w & 1-u
\end{pmatrix}
. \label{sqdm}
\end{equation}
For brevity, we will further omit the symbol of the reference
basis in notation. Of course, the real parameter $u$ lies between
$0$ and $1$. The eigenvalues of (\ref{sqdm}) are expressed as
\begin{equation}
\lambda_{\pm}=\frac{1}{2}\pm\sqrt{\left(u-\frac{1}{2}\right)^{\!2}+|w|^{2}}
\, . \label{lampm}
\end{equation}
They should be both positive and no greater than $1$, whence
$|w|\leq\sqrt{u(1-u)}$. For integer values of $\alpha$,
sufficiently simple expressions take place. For $\alpha=1$, we
obtain
\begin{equation}
\rmc_{1}(\bom)=h(u)-h(\lambda_{+})
\, , \label{hhul}
\end{equation}
where $h(u):=-\,u\ln{u}-(1-u)\ln(1-u)$ is the binary Shannon
entropy. Furthermore, we have
\begin{equation}
\rmc_{2}(\bom)=
\left(
\sqrt{u^{2}+|w|^{2}}+
\sqrt{(1-u)^{2}+|w|^{2}}
\,\right)^{\!2}-1
\, . \label{c2qb}
\end{equation}
Note also that $\rmc_{\ell_{1}}(\bom)=2\,|w|$ and
$\rmc_{\ell_{2}}(\bom)=2\,|w|^{2}$. For another integer $\alpha$,
the resulting expressions are obtained similarly to (\ref{c2qb}).

One way to study coherence of a single qubit is posed as follows.
For the given $u$, we consider an interval of changes of the corresponding quantifier. The minimum is clearly $0$, whereas the
maximum is found as for the function of $|w|\leq\sqrt{u(1-u)}$.
For example, we have
\begin{align}
\max\left\{\rmc_{2}(\bom):{\>}|w|\leq\sqrt{u(1-u)}\right\}&=2\sqrt{u(1-u)}
\, , \label{c2mu}\\
\max\left\{\rmc_{\ell_{1}}(\bom):{\>}|w|\leq\sqrt{u(1-u)}\right\}&=2\sqrt{u(1-u)}
\, . \label{cl1mu}
\end{align}
For the given $u$, the coherence quantifiers $\rmc_{2}(\bom)$ and
$\rmc_{\ell_{1}}(\bom)$ cover the same interval of values. Their
maximal values are reached for the same states. These states are
pure, since $|w|^{2}=u(1-u)$ implies $\lambda_{+}=1$ and
$\lambda_{-}=0$. The coherence quantifiers are zero for incoherent
states, when $w=0$.

\begin{figure}
\includegraphics[width=7.0cm]{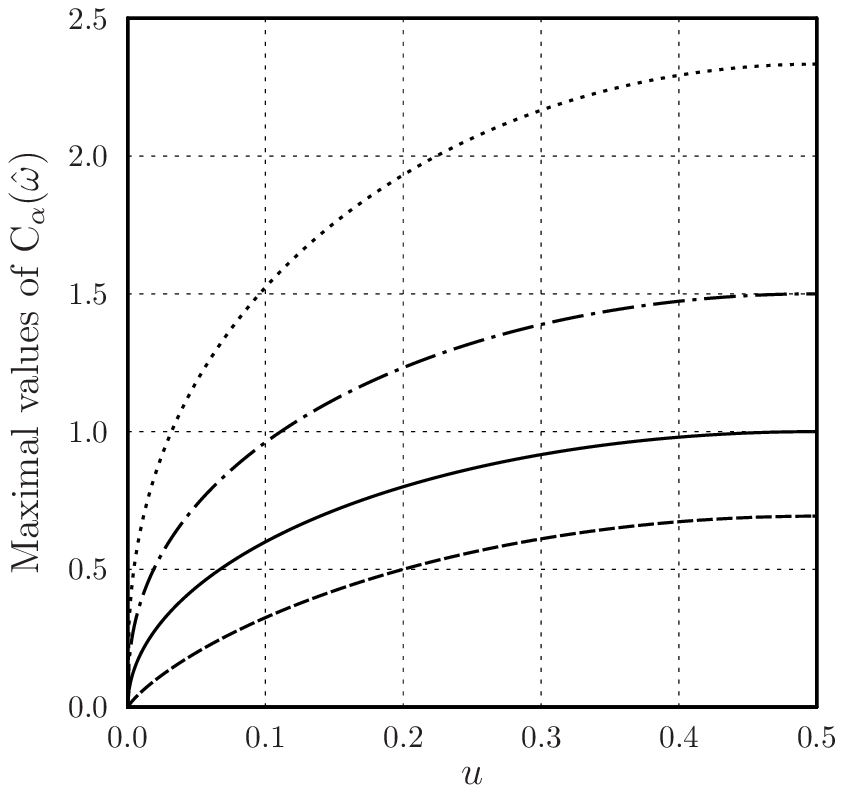}
\caption{\label{fig1}Maximal values of $\rmc_{\alpha}(\bom)$
versus $u$ are shown by a dashed line for $\alpha=1$, by a solid line
for $\alpha=2$, by a dash-dotted line for $\alpha=3$, and by a dotted
line for $\alpha=4$.}
\end{figure}

Let us consider the maximum of $\rmc_{\alpha}(\bom)$ for the given
$u$ similarly to (\ref{c2mu}) and (\ref{cl1mu}). In Fig.
\ref{fig1}, this maximum is shown as a function of $u$ for several
integer values of $\alpha$. In particular, by the solid line we
represent both (\ref{c2mu}) and (\ref{cl1mu}). The curves are
shown only for the half of the interval $u\in[0;1]$, since they
are symmetric with respect to the line $u=1/2$. The four curves
all show a similar behavior. In each case, the range between the
abscissa and the curve shows those values that are covered by the
corresponding quantifier. Hence, the coherence $\alpha$-quantifier
seems to be more sensitive for larger values of $\alpha$. On the
other hand, coherence measures should obey monotonicity
properties. Without them, any candidate to quantify coherence
cannot be accepted.

\begin{figure}
\includegraphics[width=7.0cm]{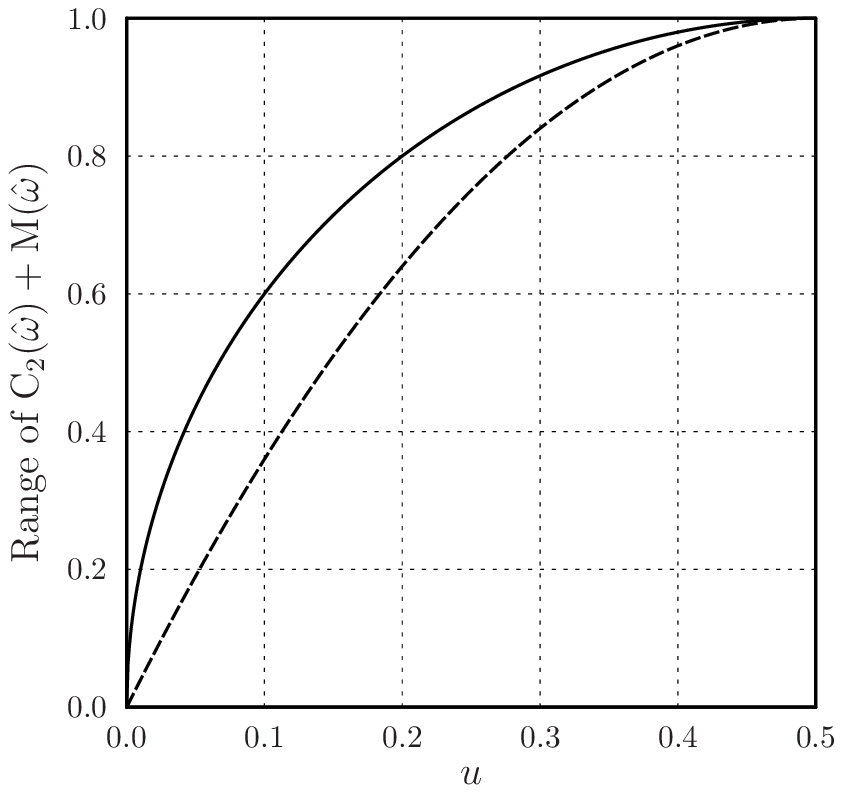}
\caption{\label{fig2}The minimal and maximal values of the sum
$\rmc_{2}(\bom)+\rmm(\bom)$ as functions of $u$.}
\end{figure}

For $d=2$, the formula (\ref{mixcr}) gives
$\rmc_{\alpha}(\bom)+\rmm(\bom)\leq1$ with $0<\alpha\leq2$. For a
single qubit, we can obtain more precise trade-off bounds.
In particular, we have
\begin{equation}
4u(1-u)\leq\rmc_{2}(\bom)+\rmm(\bom)\leq2\sqrt{u(1-u)}
\, . \label{42com}
\end{equation}
It is instructive to compare (\ref{42com}) with the exact equality
\begin{equation}
\rmc_{\ell_{1}}(\bom)^{2}+\rmm(\bom)=4u(1-u)
\, . \label{12com}
\end{equation}
In Fig. \ref{fig2}, we plot the left-hand side of (\ref{42com}) by a
dashed line and the right-hand side of (\ref{42com}) by a solid
line. Here, the former is reached for incoherent states and the
latter is reached for pure states. For each $u$, the range between
these lines shows values that are covered by the sum
$\rmc_{2}(\bom)+\rmm(\bom)$. So, this sum ranges in a narrow
interval. It is similar to the sum (\ref{12com}), but the latter is quadratic in the coherence measure. Thus, a lack of coherence will rather be accompanied
by some increasing of the mixedness. Figure \ref{fig2} also
illustrates that the relation
$\rmc_{2}(\bom)+\rmm(\bom)\leq1$ is sufficiently tight when
the diagonal elements of (\ref{sqdm}) do not differ essentially.

In the case of a single qubit, the considered coherence
quantifiers enjoy a behavior similarly to the measure
(\ref{c1nrm}). There are also natural trade-off relations between
coherence and mixedness. They are brightly exposed with the
quantifier (\ref{cma2}). It seems that the quadratic measure
(\ref{cma2}) provides a useful alternate approach to quantify
coherence. To support this claim, the question of monotonicity
should be resolved. We address this in the next section.

\section{Formulation of the monotonicity property}\label{sec6}

Desired properties of coherence measures concern their behavior
with respect to state transformations \cite{bcp14}. It is natural
to demand that coherence quantifiers cannot increase under mixing
\cite{bcp14}. Let $\{\bro_{n}\}$ be a collection of density
matrices, and let positive numbers $p_{n}$ obey $\sum_{n}p_{n}=1$.
For all $\alpha\in(0;2]$, we have
\begin{equation}
\rmc_{\alpha}\Bigl(\cle\,\Big|\sum\nolimits_{n}p_{n}\bro_{n}\Bigr)
\leq\sum\nolimits_{n}p_{n}\,\rmc_{\alpha}(\cle|\bro_{n})
\, . \label{conmixa}
\end{equation}
This result immediately follows from the the joint convexity
(\ref{joico}) and the definition (\ref{c1dt}). We refrain from
presenting the details here. For $\alpha\in(0;2]$, therefore, the
quantity (\ref{c1dt}) fulfills one of the properties listed in
\cite{bcp14}. Changes of coherence measures under some forms of
quantum operations are of great importance \cite{bcp14}. Here, the
following two classes of incoherent operations should be
considered. The first form of monotonicity property is posed as
follows. The notion of coherence is basis dependent. Let
$\cle^{\prime}$ be the prescribed orthonormal basis with respect
to which incoherent states are defined in $\hh_{B}$. We define
incoherent quantum operation as a TPCP map
$\Phi_{I}:{\>}\lnp(\hh_{A})\rightarrow\lnp(\hh_{B})$ such that its
Kraus operators all obey the property
\begin{equation}
\bro\in\cli(\cle){\>\,}\Longrightarrow{\>}
\frac{\km_{n}\bro\km_{n}^{\dagger}}{\Tr(\km_{n}\bro\km_{n}^{\dagger})}
\in\cli(\cle^{\prime})
\, . \label{eepr}
\end{equation}
For $\alpha\in(0;2]$, the coherence quantifier (\ref{c1dt}) is
monotone under incoherent quantum operations, namely,
\begin{equation}
\rmc_{\alpha}\bigl(\cle^{\prime}|\Phi_{I}(\bro)\bigr)\leq\rmc_{\alpha}(\cle|\bro)
\, .  \label{prp2}
\end{equation}
This follows from the property (\ref{mnren}) and the definition
(\ref{c1dt}), which includes the minimization.

Monotonicity under incoherent selective measurements seems to be
more sophisticated \cite{bcp14}. Formulating (\ref{prp2}), we
assume the loss of information about the measurement outcome. When
measurement outcomes are retained, one further allows a
subselection according to these outcomes. Such operations are also
described by a set of Kraus operators $\{\km_{n}\}$, but now these
operators may have different output spaces though the input space
is the same. So, we consider a set of operators
$\km_{n}:{\>}\hh_{A}\to\hh_{Bn}$ that satisfy (\ref{clrl}). To
each output space $\hh_{Bn}$, we assign the orthonormal basis
$\cle_{n}^{\prime}$ used for determining incoherent density
matrices. The authors of \cite{bcp14} formulated the monotonicity
under incoherent selective measurements as
\begin{equation}
\sum\nolimits_{n}p_{n}\,\rmc\bigl(\cle_{n}^{\prime}\big|\bro_{n}\bigr)
\leq\rmc(\cle|\bro)
\, . \label{consel}
\end{equation}
Here, $p_{n}=\Tr\bigl(\km_{n}\bro\km_{n}^{\dagger}\bigr)$ is the
probability of $n$-th outcome resulting in $n$-th particular
output
\begin{equation}
\bro_{n}=p_{n}^{-1}\km_{n}\bro\km_{n}^{\dagger}
\, . \label{paout}
\end{equation}
The authors of \cite{rpl15} called (\ref{consel}) the
strong monotonicity under incoherent channels. In the context of
transport phenomena, incoherent quantum channels are considered in
\cite{lm14}. It turns out that the coherence $\alpha$-quantifiers
obey the monotonicity property in the following form.

\newtheorem{prop43}[prop31]{Theorem}
\begin{prop43}\label{thm43}
Let the incoherent state $\bdl_{\rho}\in\cli(\cle)$ be such that
$\rmc_{\alpha}(\cle|\bro)=\rmd_{\alpha}(\bro||\bdl_{\rho})$. For
all $\alpha\in(0;2]$, the coherence measures (\ref{c1dt}) are
changed under any incoherent selective measurement in line with
\begin{equation}
\sum\nolimits_{n}p_{n}^{\alpha}q_{n}^{1-\alpha}\,\rmc_{\alpha}(\cle^{\prime}_{n}|\bro_{n})
\leq\rmc_{\alpha}(\cle|\bro)
\, , \label{insela}
\end{equation}
where $p_{n}=\Tr(\km_{n}\bro\km_{n}^{\dagger})$,
$q_{n}=\Tr(\km_{n}\bdl_{\rho}\km_{n}^{\dagger})$, and
$\bro_{n}$ is defined by (\ref{paout}).
\end{prop43}

{\bf Proof.} We first note that we can consider the set of Kraus
operators with the same output space. More precisely, we define
the space
\begin{equation}
\widetilde{\hh}_{B}:=
\bigoplus_{n}\hh_{Bn}
\, , \label{samout}
\end{equation}
where $\hh_{Bn}$ is the output space of the $n$-th Kraus
operator $\km_{n}$. To each $\km_{n}:{\>}\hh_{A}\to\hh_{Bn}$, we
assign the operator
\begin{equation}
\wkm_{n}=
\begin{pmatrix}
\niz \\
\cdots \\
\km_{n} \\
\cdots \\
\niz
\end{pmatrix}
. \label{wkmn0}
\end{equation}
That is, this operator is represented as a block column, whose
$n$-th block is $\km_{n}$ and others are all zero. The operator
$\wkm_{n}$ maps vectors of $\hh_{A}$ to vectors of
$\widetilde{\hh}_{B}$. For each $n$, the input states $\bro$ and
$\bdl_{\rho}$ are respectively mapped into subnormalized outputs
\begin{align}
\wkm_{n}\bro\wkm_{n}^{\dagger}
&={\mathrm{diag}}\bigl(\niz\cdots\km_{n}\bro\km_{n}^{\dagger}\cdots\niz\bigr)
, \label{wkmn1}\\
\wkm_{n}\bdl_{\rho}\wkm_{n}^{\dagger}
&={\mathrm{diag}}\bigl(\niz\cdots\km_{n}\bdl_{\rho}\km_{n}^{\dagger}\cdots\niz\bigr)
. \label{wkmn2}
\end{align}
Thus, the output (\ref{wkmn1}) is the diagonal block matrix with
the $(n,n)$-block $\km_{n}\bro\km_{n}^{\dagger}$ and other zero
blocks. Similarly, the output (\ref{wkmn2}) is the diagonal block
matrix with the $(n,n)$-block
$\km_{n}\bdl_{\rho}\km_{n}^{\dagger}$. According to the definition
(\ref{qendf}), we then have
\begin{equation}
\rmd_{\alpha}
\bigl(
\wkm_{n}\bro\wkm_{n}^{\dagger}\big|\big|\wkm_{n}\bdl_{\rho}\wkm_{n}^{\dagger}
\bigr)
=\rmd_{\alpha}
\bigl(
\km_{n}\bro\km_{n}^{\dagger}\big|\big|\km_{n}\bdl_{\rho}\km_{n}^{\dagger}
\bigr)
. \label{wweqs}
\end{equation}
We define a TPCP map
$\widetilde{\Phi}_{I}:{\>}\lnp(\hh_{A})\rightarrow\lnp\bigl(\widetilde{\hh}_{B}\bigr)$
by the formula
\begin{equation}
\widetilde{\Phi}_{I}(\ax):=
\sum\nolimits_{n}\wkm_{n}\ax\wkm_{n}^{\dagger}
\, . \label{wphdf0}
\end{equation}
It is trace preserving due to
$\wkm_{n}^{\dagger}\wkm_{n}=\km_{n}^{\dagger}\km_{n}$
and the fact that Kraus operators of any incoherent selective
measurement obey (\ref{clrl}). With the input state $\bro$, the
output of the quantum operation (\ref{wphdf0}) can be represented
as
\begin{equation}
\widetilde{\Phi}_{I}(\bro)=\sum\nolimits_{n}
p_{n}\wbro_{n}
\, , \label{wphdf1}
\end{equation}
where
$\wbro_{n}=p_{n}^{-1}\wkm_{n}\bro\wkm_{n}^{\dagger}$.
Denoting
$\bdl_{\rho{n}}=q_{n}^{-1}\km_{n}\bdl_{\rho}\km_{n}^{\dagger}$, we
write the following relations:
\begin{align}
\rmd_{\alpha}(\bro||\bdl_{\rho})&\geq
\rmd_{\alpha}\bigl(\widetilde{\Phi}_{I}(\bro)\big|\big|\widetilde{\Phi}_{I}(\bdl_{\rho})\bigr)
\label{comon1}\\
&=\sum\nolimits_{n}\rmd_{\alpha}
\bigl(\wkm_{n}\bro\wkm_{n}^{\dagger}
\big|\big|\wkm_{n}\bdl_{\rho}\wkm_{n}^{\dagger}\bigr)
\label{comon2}\\
&=\sum\nolimits_{n}\rmd_{\alpha}
\bigl(\km_{n}\bro\km_{n}^{\dagger}\big|\big|\km_{n}\bdl_{\rho}\km_{n}^{\dagger}\bigr)
\label{comon3}\\
&\geq
\sum\nolimits_{n}p_{n}^{\alpha}q_{n}^{1-\alpha}\,
\rmd_{\alpha}(\bro_{n}||\bdl_{\rho{n}})
\, . \label{comon4}
\end{align}
Here, step (\ref{comon1}) follows from (\ref{mnren}), and step
(\ref{comon2}) follows from (\ref{tdit}). Indeed, the construction
of operators (\ref{wkmn0}) implies orthogonality of subspaces
$\ron(\wkm_{n}\bro\wkm_{n}^{\dagger})$ for
different indices $n$. Further, step (\ref{comon3}) follows
from (\ref{wweqs}), and step (\ref{comon4}) is based on Theorem
\ref{thm32}. The inequality
$\rmd_{\alpha}(\bro_{n}||\bdl_{\rho{n}})\geq\rmc_{\alpha}(\cle^{\prime}_{n}|\bro_{n})$,
clear from definition (\ref{c1dt}), completes the proof.
$\blacksquare$

In the case $\alpha=1$, the statement of Theorem \ref{thm43}
reduces to (\ref{consel}) written with the coherence measure
$\rmc_{1}(\cle|\bro)$. This property was first proved in
\cite{bcp14}. In a certain sense, the relation (\ref{insela}) is a
natural extension of the formula (\ref{consel}). The coherence
$\alpha$-measures can also be treated as monotone, but the
inequality is posed formally in a more sophisticated manner. In
particular, the formulation now involves the particular probabilities
$q_{n}$ calculated for the incoherent state (\ref{bdlra}). Thus,
the probabilities $q_{n}$ are also dependent on the considered
state $\bro$, but not so directly as $p_{n}$'s. In view of the
above results, constructing coherence measures that obey
monotonicity just in the form (\ref{consel}) seems to be
difficult. The only known examples are the measures
$\rmc_{\ell_{1}}(\cle|\bro)$ and $\rmc_{1}(\cle|\bro)$. Of course,
we do not consider here any linear combination of the mentioned
two measures. As the properties imposed are linear in a coherence
measure, a linear combination of two (or more) particular measures
will obey these properties whenever each particular measure does.

\section{Two quadratic measures compared}\label{sec7}

In this section, we will compare two quantifiers of coherence obtained
as homogeneous quadratic functions of matrix elements. These
measures are respectively defined by the formulas (\ref{c22nrm})
and (\ref{cma2}). The authors of \cite{bcp14} exemplified that the
coherence measure (\ref{c22nrm}) is not monotone under incoherent
selective measurements. It is instructive to examine the property
(\ref{insela}) just with this example. Let us check monotonicity
of the coherence measure (\ref{cma2}). In the example considered,
the input and output reference bases are the same. For brevity, we
will omit the symbols $\cle$ and $\cle^{\prime}$ in further
calculations. The two Kraus operators are written as
\begin{equation}
\km_{1}=
\begin{pmatrix}
0 & 1 & 0 \\
0 & 0 & 0 \\
0 & 0 & a
\end{pmatrix}
, \qquad
\km_{2}=
\begin{pmatrix}
1 & 0 & 0 \\
0 & 0 & b \\
0 & 0 & 0
\end{pmatrix}
, \label{kk12}
\end{equation}
where the complex numbers $a$ and $b$ obey $|a|^{2}+|b|^{2}=1$.
Further, one considers the density matrix
\begin{equation}
\vbro=
\frac{1}{4}
\begin{pmatrix}
1 & 0 & 1 \\
0 & 2 & 0 \\
1 & 0 & 1
\end{pmatrix}
. \label{php21}
\end{equation}
The normalized particular outputs are expressed as
\begin{align}
&\vbro_{1}=
\frac{1}{2+|a|^{2}}
\begin{pmatrix}
2 & 0 & 0 \\
0 & 0 & 0 \\
0 & 0 & |a|^{2}
\end{pmatrix}
, \label{php120}\\
&\vbro_{2}=
\frac{1}{1+|b|^{2}}
\begin{pmatrix}
1 & b^{*} & 0 \\
b & |b|^{2} & 0 \\
0 & 0 & 0
\end{pmatrix}
. \label{php12}
\end{align}
The corresponding probabilities are written as
\begin{equation}
p_{1}=\frac{2+|a|^{2}}{4}
\ , \qquad
p_{2}=\frac{1+|b|^{2}}{4}
\ . \label{prb12}
\end{equation}
For $\alpha=2$, the $\alpha$-divergence is minimized with the
incoherent state, whose non-zero entries are proportional to the
square roots of the diagonal elements of $\vbro^{2}$:
\begin{equation}
\bdl_{\varrho}=
\frac{1}{2+\sqrt{2}}
\begin{pmatrix}
1 & 0 & 0 \\
0 & \sqrt{2} & 0 \\
0 & 0 & 1
\end{pmatrix}
. \label{bd21}
\end{equation}
Calculations of the coherence measure (\ref{cma2}) result in
\begin{equation}
\rmc_{2}(\vbro)=\rmd_{2}(\vbro||\bdl_{\varrho})=\frac{2\sqrt{2}-1}{4}
\ . \label{c2vb}
\end{equation}
Using $\bdl_{\varrho}$ as the input, we get the probabilities of
particular outcomes,
\begin{align}
&q_{1}=\Tr(\km_{1}\bdl_{\varrho}\km_{1}^{\dagger})=\frac{\sqrt{2}+|a|^{2}}{2+\sqrt{2}}
\ , \label{q1q1}\\
&q_{2}=\Tr(\km_{2}\bdl_{\varrho}\km_{2}^{\dagger})=\frac{1+|b|^{2}}{2+\sqrt{2}}
\ . \label{q1q2}
\end{align}
For the density matrices (\ref{php12}), we obtain
$\rmc_{2}(\vbro_{1})=0$ and
\begin{equation}
\rmc_{2}(\vbro_{2})=\frac{2\,|b|}{1+|b|^{2}}
\ . \label{lbc2}
\end{equation}
For all $|b|\in[0;1]$, we consider the quantity
\begin{align}
&p_{1}^{2}q_{1}^{-1}\rmc_{2}(\vbro_{1})+
p_{2}^{2}q_{2}^{-1}\rmc_{2}(\vbro_{2})
=\frac{2+\sqrt{2}}{8}\,|b|
\nonumber\\
&\leq\frac{2+\sqrt{2}}{8}
\approx0.4268
\, , \label{p1q2}
\end{align}
which is strictly less than
$\rmc_{2}(\vbro)=\bigl(2\sqrt{2}-1\bigr)/4\approx0.4571$. The
latter point illustrates the result (\ref{insela}). The example
also shows that the formulation (\ref{insela}) is actually
necessary. Indeed, the quantifier (\ref{cma2}) does not
share the monotonicity formulation (\ref{consel}). To see this
fact, we write
\begin{equation}
p_{1}\,\rmc_{2}(\vbro_{1})+
p_{2}\,\rmc_{2}(\vbro_{2})
=\frac{|b|}{2}
\ . \label{q2p1}
\end{equation}
The right-hand side of (\ref{q2p1}) increases up to $0.5$ for
$|b|=1$ and can exceed $\rmc_{2}(\vbro)\approx0.4571$. The above
findings give an evidence that the $\alpha$-quantifiers do not
generally obey monotonicity in the form of (\ref{consel}). On the
other hand, these measures certainly satisfy monotonicity in the
form of (\ref{insela}). Thus, the monotonicity of coherence under
selective measurements is sophisticated in character. This
property does not follow immediately from the monotonicity of
quantum relative entropies.

The considered example allows us to resolve the following natural
question. The coherence measure (\ref{c22nrm}) does not share
monotonicity in the form of (\ref{consel}). In principle, we may ask
for monotonicity of (\ref{c22nrm}) similarly to (\ref{insela}). In
other words, we consider the quantity
\begin{equation}
\sum\nolimits_{n}p_{n}^{\alpha}r_{n}^{1-\alpha}\,\rmc_{\ell_{2}}(\bro_{n})
\, , \label{insela2}
\end{equation}
where $r_{n}=\Tr(\km_{n}\bdl_{*}\km_{n}^{\dagger})$ and $\bdl_{*}$
is obtained from $\bro$ by vanishing all off-diagonal entries. The
above example shows that (\ref{insela2}) can exceed
$\rmc_{\ell_{2}}(\bro)$. Indeed, the input state (\ref{php21}) is
such that $r_{n}=p_{n}$ for $n=1,2$. Hence, the quantity
(\ref{insela2}) is equal to
$\sum_{n}p_{n}\,\rmc_{\ell_{2}}(\vbro_{n})$ and violates
monotonicity in the form of (\ref{consel}), as already known. Thus,
the coherence measure (\ref{c22nrm}) is not monotone even in the
sense of (\ref{insela}). This fact shows that the formulation
(\ref{insela}) is not trivial. It is also reduced to
(\ref{consel}) in the limit $\alpha\to1$. Thus, we can treat
(\ref{insela}) as a natural extension of the standard form
(\ref{consel}).

\section{Conclusions}\label{sec8}

We have examined quantum-coherence measures based on
$\alpha$-divergences of the Tsallis type. Trade-off relations
between coherence and mixedness were obtained. Some properties
were further exemplified with a single qubit. Most of the desired
properties immediately follows from general properties of quantum
relative entropies. The monotonicity of coherence under selective
measurements is a more interesting and complicated question. This
monotonicity has been shown for the two measures based on the
$\ell_{1}$-norm and on the standard relative entropy \cite{bcp14}.
For the coherence measure based on the trace distance, only
particular monotonicity results are known \cite{rpl15}. We have
proved that coherence $\alpha$-measures enjoy desired monotonicity
in the form of (\ref{insela}), where the parameter $\alpha$ is
involved. For $\alpha\to1$, this formulation is directly reduced
to the standard formulation proposed in \cite{bcp14}. In this
regard, the result (\ref{insela}) is a parametric extension of the
standard form (\ref{consel}). It may be supposed that the two
known examples satisfying just (\ref{consel}) are the only such.

The obtained family includes the quantity expressed in terms of
the squared moduli of matrix elements. In several respects, this
quantity differs from the coherence measure induced by the squared
$\ell_{2}$-norm. In both (\ref{c1nrm}) and (\ref{c22nrm}), the
closest incoherent state is obtained by vanishing all off-diagonal
entries of $\bro$. Except for $\alpha=1$, the incoherent state
that minimizes the $\alpha$-divergence in (\ref{c1dt}) is reached
by a more complicated procedure. Nevertheless, for quantum
$\alpha$-divergences of the Tsallis type, the required minimization
can be solved with an explicit answer. It seems to be difficult for
quantum $f$-divergences in general. Currently, so-called
``sandwiched'' relative entropies are the subject of active
research \cite{mdsft13}. Such quantities could be used for
obtaining of coherence measures but he required minimization seems
to be difficult. By comparing two quadratic measures of coherence, we
also have shown that the measure induced by the squared $\ell_{2}$-norm
violates monotonicity even in an generalized form. It was
conjectured in \cite{hall15} that the square root of
(\ref{c22nrm}) may obey all the desired properties. Due to our
results, this conjecture seems to be sufficiently difficult to
resolve.


\bigskip

\begin{thebibliography}{50}

\bibitem{mandelw}
L.~Mandel and E.~Wolf, {\it Optical Coherence and Quantum Optics} (Cambridge University Press, Cambridge 1995).

\bibitem{horodecki13}
M.~Horodecki and J.~Oppenheim, Nat. Commun. {\bf 4}, 2059 (2013).

\bibitem{rfag13}
C.~Rodr\'{i}guez-Rosario, T.~Frauenheim, and A.~Aspuru-Guzik, arXiv:1308.1245 [quant-ph] (2013).

\bibitem{rudolph15}
M.~Lostaglio, D.~Jennings, and T.~Rudolph, Nat. Commun. {\bf 6}, 6383 (2015).

\bibitem{rudolx15}
M.~Lostaglio, K.~Korzekwa, D.~Jennings, and T.~Rudolph, Phys. Rev. X {\bf 5}, 021001 (2015).

\bibitem{ngour15}
V.~Narasimhachar and G.~Gour, Nat. Commun. {\bf 6}, 7689 (2015).

\bibitem{nielsen}
M.~A.~Nielsen and I.~L.~Chuang, {\it Quantum Computation and Quantum Information} (Cambridge University Press, Cambridge, 2000).

\bibitem{spekkens08}
G.~Gour and R.~W.~Spekkens, New J. Phys. {\bf 10}, 033023 (2008).

\bibitem{bhors13}
F.~G.~S.~L.~Brand\~{a}o, M.~Horodecki, J.~Oppenheim, J.~M.~Renes,
and R.~W.~Spekkens, Phys. Rev. Lett. {\bf 111}, 250404 (2013).

\bibitem{bcp14}
T.~Baumgratz, M.~Cramer, and M.~B.~Plenio, Phys. Rev. Lett. {\bf 113}, 140401 (2014).

\bibitem{winter15}
A.~Winter and D.~Yang, arXiv:1506.07975 [quant-ph] (2015).

\bibitem{hall15}
S.~Cheng and M.~J.~W.~Hall, Phys. Rev. A {\bf 92}, 042101 (2015).

\bibitem{mondal15}
D.~Mondal, T.~Pramanik, and A.~K.~Pati, arXiv:1508.03770 [quant-ph] (2015).

\bibitem{bera15}
M.~N.~Bera, T.~Qureshi, M.~A.~Siddiqui, and A.~K.~Pati, Phys. Rev. A {\bf 92}, 012118 (2015).

\bibitem{bagan15}
E.~Bagan, J.~A.~Bergou, S.~S.~Cottrell, and M.~Hillery, arXiv:1509.04592 [quant-ph] (2015).

\bibitem{hillery15}
M.~Hillery, Phys. Rev. A {\bf 93}, 012111 (2016).

\bibitem{bca15}
T.~R.~Bromley, M.~Cianciaruso, and G.~Adesso, Phys. Rev. Lett. {\bf 114}, 210401 (2015).

\bibitem{vedral15}
B.~Yadin, J.~Ma, D.~Girolami, M.~Gu, and V.~Vedral, arXiv:1512.02085 [quant-ph] (2015).

\bibitem{jxu15}
J.~Xu, Phys. Rev. A {\bf 93}, 032111 (2016).

\bibitem{zslf16}
Y.-R.~Zhang, L.-H.~Shao, Y.~Li, and H.~Fan, Phys. Rev. A {\bf 93}, 012334 (2016).

\bibitem{hornJ}
R.~A.~Horn and C.~R.~Johnson, {\it Matrix Analysis} (Cambridge University Press, Cambridge, 1985).

\bibitem{rpl15}
S.~Rana, P.~Parashar, and M.~Lewenstein, Phys. Rev. A {\bf 93}, 012110 (2016).

\bibitem{ssdba15}
A.~Streltsov, U.~Singh, H.~S.~Dhar, M.~N.~Bera, and G.~Adesso, Phys. Rev. Lett. {\bf 115}, 020403 (2015).

\bibitem{uhlmann76}
A.~Uhlmann, Rep. Math. Phys. {\bf 9}, 273 (1976).

\bibitem{jozsa94}
R.~Jozsa, J. Mod. Opt. {\bf 41}, 2315 (1994).

\bibitem{bengtsson}
I.~Bengtsson and K.~\.{Z}yczkowski, {\it Geometry of Quantum
States: An Introduction to Quantum Entanglement} (Cambridge
University Press, Cambridge, 2006).

\bibitem{hmpb11}
F.~Hiai, M.~Mosonyi, D.~Petz, and C.~B\'{e}ny, Rev. Math. Phys. {\bf 23}, 691 (2011).

\bibitem{ics67}
I.~Csisz\'{a}r, Studia Sci. Math. Hungar. {\bf 2}, 299 (1967).

\bibitem{borland}
L.~Borland, A.~R.~Plastino, and C.~Tsallis, J. Math. Phys. {\bf 39}, 6490 (1998).

\bibitem{fky04}
S.~Furuichi, K.~Yanagi, and K.~Kuriyama, J. Math. Phys. {\bf 45}, 4868 (2004).

\bibitem{vajda06}
F.~Liese and I.~Vajda, IEEE Trans. Inf. Theor. {\bf 52}, 4394 (2006).

\bibitem{vedral02}
V.~Vedral, Rev. Mod. Phys. {\bf 74}, 197 (2002).

\bibitem{petz86}
D.~Petz, Rep. Math. Phys. {\bf 23}, 57 (1986).

\bibitem{bhatia07}
R.~Bhatia, {\it Positive Definite Matrices} (Princeton University Press, Princeton, 2007).

\bibitem{vepl98}
V.~Vedral and M.~B.~Plenio, Phys. Rev. A {\bf 57}, 1619 (1998).

\bibitem{rast13b}
A.~E.~Rastegin, Eur. Phys. J. D {\bf 67}, 269 (2013).

\bibitem{brz99}
\u{C}.~Brukner and A.~Zeilinger, Phys. Rev. Lett. {\bf 83}, 3354 (1999).

\bibitem{bz10}
T.~Durt, B.-G.~Englert, I.~Bengtsson, and K.~\.{Z}yczkowski, Int. J. Quantum Inf. {\bf 08}, 535 (2010).

\bibitem{rastproca}
A.~E.~Rastegin, Proc. Roy. Soc. A {\bf 471}, 20150435 (2015).

\bibitem{kwiat2004}
N.~A.~Peters, T.-C.~Wei, and P.~G.~Kwiat, Phys. Rev. A {\bf 70}, 052309 (2004).

\bibitem{sbdp15}
U.~Singh, M.~N.~Bera, H.~S.~Dhar, and A.~K.~Pati, Phys. Rev. A {\bf 91}, 052115 (2015).

\bibitem{lm14}
F.~Levi and F.~Mintert, New J. Phys. {\bf 16}, 033007 (2014).

\bibitem{mdsft13}
M.~M\"{u}ller-Lennert, F.~Dupuis, O.~Szehr, S.~Fehr, and
M.~Tomamichel, J. Math. Phys. {\bf 54}, 122203 (2013).


\end{thebibliography}
\end{document}